\documentclass[a4paper,12pt]{article}
\usepackage{graphicx}
\usepackage{cite}
\usepackage{amssymb,amsmath,latexsym,enumerate}
\usepackage{subfigure,epsfig}
\usepackage{array,longtable,lscape,bigints}
\usepackage{color}

\begin{document}

\title{Integrable model of nonlinear dislocations}

\author{Nikolay A. Kudryashov}

\date{Department of Applied Mathematics, National Research Nuclear University MEPhI, 31 Kashirskoe Shosse, 115409 Moscow, Russian Federation}

\maketitle

\begin{abstract}

Using the continuous limit approximation in the dynamical system we study a nonlinear partial differential equation which corresponds to the generalization of both the Fermi-Pasta-Ulam and the Frenkel-Kontorova models. This generalized model can be considered as a model for the description nonlinear dislocation waves in the crystal lattice. We obtain the nonlinear partial differential equation for the description of dislocations. Taking into account the wave moving in one direction we obtain another nonlinear evolution equation. Using the Painlev\'e test we analyze the integrability of this equation. We find that there exist an integrable case of the partial differential equation for nonlinear dislocations. The Lax pair for the solution of the Cauchy problem is found. The solution of the Cauchy problem for nonlinear evolution equation is discussed. One and the two-soliton solutions for the nonlinear evolution equation are presented. The influence of parameters on the propagation of one and the two-soliton solutions is analyzed and demonstrated.

\end{abstract}

\section{Introduction}

Let us consider the following dynamical system:
\begin{equation}\begin{gathered}
\label{I1}
m\,\frac{d^2y_{i}}{d t}=\left(y_{i+1}-2\,y_i+y_{i-1}\right)\,
\left[k+ \alpha\,(y_{i+1}-y_{i-1})+\right.\\
\\
\left.+\beta\,(y_{i+1}^2+y_i^2+y_{i-1}^2-y_{i+1}y_i-y_{i+1} y_{i-1}-y_i y_{i-1})\right]-f_0\,\sin{\left(\frac{2\,\pi\,y_i}{a}\right)}, \\
\\
(i=1,\ldots,N),
\end{gathered}\end{equation}
where $y_i$ denotes the displacement of the i-th mass from its original position, $ t$ is time,
$k$, $\alpha$, $\beta$, $f_0$ and $a$ are constant parameters of system \eqref{I1}.

The system of equations \eqref{I1} is the generalization of some well-known dynamical systems. At $\alpha=0$ and $\beta=0$ the system of equations \eqref{I1} is the mathematical model by Frenkel and Kontorova for the description of dislocations in the rigid body \cite{Frenkel, Frenkelbook, Braun}. It was suggested in this model that the influence of atoms in the crystal is taken into account by term $f_0\,\sin {\frac{2\,\pi\,y_i}{a}}$ but the dislocations interact by means of linear low. Assuming that $N\rightarrow \infty$ and $h\rightarrow 0$ where $h$ is the distance between atoms, one can get the Sine-Gordon equation \cite{Braun}
\begin{equation}\begin{gathered}
\label{I2}
u_{x\tau}=\sin{u}
\end{gathered}\end{equation}

In the case of $f=0$ and $\beta=0$ the system of equations \eqref{I1} is the well-known Fermi-Pasta-Ulam model \cite{Fermi} which was studied many times \cite{Porter, Ruffo, Genta, Berman, Ford}. It is known that the Fermi-Pasta-Ulam model $N\rightarrow \infty$ and $h\rightarrow 0$ is transformed to the Korteweg-de Vries equation \cite{Kruskal}
\begin{equation}\begin{gathered}
\label{I3}
u_{\tau}+6\,u\,u_{x}+u_{xxx}=0.
\end{gathered}\end{equation}

The main result of work \cite{Kruskal} was the introduction of solitons which are solutions of equation \eqref{I3}.
It was shown in 1967 that the Cauchy problem for the Korteweg-de Vries equation \eqref{I3} can be solved by the Inverse Scattering transform \cite{Gardner}. In the paper \cite{Kudr15} the author took into account high order terms in the Taylor series for the description of nonlinear waves in the $\alpha $ Fermi-Pasta-Ulam model.

Assuming $f_0=0$, $\alpha \neq 0$ and $\beta \neq 0$  at $N \rightarrow \infty$ and $h \rightarrow 0$ one can obtain the modified Korteweg-de Vries equation in the form \cite{Kudr16}
\begin{equation}\begin{gathered}
\label{I4}
v_{\tau}+6\,v^2\,v_{x}+v_{xxx}=0.
\end{gathered}\end{equation}

These facts bring up the question, whether the situation is similar for equation \eqref{I1} at $f_0 \neq 0$, $\alpha \neq 0$ and $\beta \neq 0$ when we assume that the interaction between dislocations in crystal is described by means of nonlinear low at $\alpha \neq 0 $ and $\beta \neq 0$.

The aim of the present Letter is to derive the nonlinear partial differential equation corresponding to dynamical system \eqref{I1} and to study the properties of this equation.

The rest of this work is organized as follows. In Section 2 we derive the fourth-order partial differential equation for the description of the nonlinear dislocation waves described by system \eqref{I1}. In Section 3 we apply the Painlev\'e test to study analytical properties of nonlinear equation. In Section 4 we present the Lax pair for solving the Cauchy problem for the nonlinear partial differential equation corresponding to dynamical system \eqref{I1}. In Section 5 we discuss the solution of the Cauchy problem for the special case of nonlinear partial differential equation. We analyze the one and the two-soliton solutions of nonlinear differential equation corresponding to dynamical system \eqref{I1} at  $f_0 \neq 0$, $\alpha \neq 0$ and $\beta \neq 0$ in Section 65. In Section 7 we briefly discuss the results of this work.

\section{Derivation of the model for the description of nonlinear dislocations}

Let us assume that the suggestions $N \rightarrow \infty$ and $h \rightarrow 0$ are carried out for dynamical systems \eqref{I1}. In this case we can use the continuous limit approximation in dynamical system \eqref{I1}. Taking into consideration the expansion of the deviation of mass $y_{i\pm 1}$ from equilibrium  position in the Taylor series up to $h^4$, we obtain
\begin{equation}\begin{gathered}
\label{D1}
y_{i\pm 1}=y_i\,\pm\, h\, y_{i,x} +\frac{h^2}{2}\,y_{i,xx} \, \pm\, \frac{h^3}{6}\,y_{i,xxx}\,
+\,\frac{h^4}{24}\,y_{i,xxxx}\,+\ldots.
\end{gathered}\end{equation}
Substituting expansion \eqref{D1} into equation \eqref{I1} and using the expressions with order $h^4$ inclusively, we obtain the nonlinear fourth -- order partial differential equation in the form:
\begin{equation}\begin{gathered}
\label{D2}
m\,y_{tt}=k\,{h}^{
2}\,y_{xx}+2 \alpha h^3 y_x\,  y_{xx}+3 \beta h^4 y_x^2\, y_{xx}+\frac{k h^4}{12} y_{xxxx}-f_0 \sin{\left(\frac{2 \pi y}{a}\right)}.
\end{gathered}\end{equation}
Using new parameters and a new variable we have
\begin{equation}\begin{gathered}
\label{D3}
c=\frac{h\,\sqrt{k}}{\sqrt{m}},\quad \varepsilon=\,\frac {a\,h}{\pi\,k}, \quad \beta^{'}=\,\frac {\beta\, h\,a}{4\,\pi}, \\
\\
v(x,\tau)=\frac{2\,\pi}{a}\,y(x,\tau),\, \quad
\gamma^{'}=\frac{\pi\,k\,h}{12\,a},\quad \delta=\frac{2\,\pi^2\,f_0}{a^2\,h^3}.
\end{gathered}\end{equation}
equation \eqref{D2} can be written as the following:
\begin{equation}\begin{gathered}
\label{D4}
\,v_{tt}=c^2\,v_{xx}+\alpha\,c^2\,\varepsilon\,v_x\,v_{xx}+
3\,\beta^{'}\,c^2\,\varepsilon\,v_x^2\,v_{xx}+\gamma^{'}\,c^2\,\varepsilon\,v_{xxxx}
-\delta\,c^2\,\varepsilon\,\sin {v}.
\end{gathered}\end{equation}

Using the nonlinear wave with a motion in the right hand side
\begin{equation}\begin{gathered}
\label{D4}
v(x,t)=u(x^{'},t^{'})+\varepsilon \,v_1(x,t),
\end{gathered}\end{equation}
where
\begin{equation}\begin{gathered}
\label{D4a}
x^{'}=x-c\,t^{'},\qquad t^{'}=\frac{c}{2}\,\varepsilon\,t
\end{gathered}\end{equation}
(primes are omitted) we obtain the equation in the form
\begin{equation}\begin{gathered}
\label{PDE}
u_{x t}+
\alpha \,u_x\,u_{xx}+3\,\beta\,u_x^2\,u_{xx}\,+\gamma\,u_{xxxx}=\delta\,\sin{u}.
\end{gathered}\end{equation}

The last equation describes the propagation of nonlinear waves in the crystalline lattice. This equation generalizes the Sine-Gordon equation in the case of $\beta \neq 0$ and $\gamma \neq 0$, when there is nonlinear interaction in the mass chain.  On other hand equation \eqref{PDE} is the generalization of the modified Korteweg-de Vries equation because at $\delta=0$ it can be transsformed to them. Let us study the integrability of equation \eqref{PDE} in the general case.

\section{The Painlev\'e test for equation \eqref{PDE}}

Let us apply the Painlev\'e test to study analytical properties of equation $\eqref{PDE}$. With this aim we use  the variables of traveling wave solutions for the equation in the form
\begin{equation}\begin{gathered}
\label{PT1}
u(x,t)=y(z),\qquad z=x-C_0\,t.
\end{gathered}\end{equation}
In this case equation \eqref{PDE} takes the form
\begin{equation}\begin{gathered}
\label{PT2}
\gamma\,y_{zzzz}+3\,\beta\,y_z^2\,y_{zz}+\alpha\,y_z\,y_{zz}-C_0\,y_{zz}-\delta\,\sin y=0.
\end{gathered}\end{equation}

Using the new variable $y(z)=\frac{i}{2}\,\ln(v(z)$ we obtain the algebraic form of equation \eqref{PT2}
\begin{equation}\begin{gathered}
\label{PT3}
\gamma\,v^3\,v_{zzzz}-4\,\gamma\,v^2\,v_z\,v_{zzz}+12\,\gamma\,v\,v_z^2\,v_{zz}-3\,\gamma\,v^2\,v_{zz}^2-6\,\gamma\,v_z^4-
i\,\alpha\,v^2\,v_z\,v_{zz}+\\
\\+i\,\alpha\,v\,v_z^3-3\,\beta\,v\,v_z^2\,v_{zz}+3\,\beta\,v_z^4-\frac12\,\delta\,v^5+\frac12\,\delta\,v^3-\\
\\
-C_0\,v^3\,v_{zz}+C_0\,v^2\,v_z^2=0.
\end{gathered}\end{equation}

The solution of equation \eqref{PT3} has the fourth-order pole. The first member of the expansion for solution in the Laurent series is the following
\[v =\frac{48(\gamma\,-8\,\delta)}{\delta\,z^4} + \ldots.\]
Using the second step of the Painlev\'e test \cite{Ramani}, we obtain the Fuchs indices in the form
\begin{equation}\begin{gathered}
\label{PT4}
j_1=-1,\quad j_2=4,\quad j_{3,4}=\frac{3\,\gamma\pm \sqrt{192\,\beta\,\gamma-15\,\gamma^2}}{2\,\gamma}.
\end{gathered}\end{equation}

We obtain that all the Fuchs indices are integers in the case
\begin{equation}\begin{gathered}
\label{PT5}
\gamma=\frac{48\,\beta}{N^2-3\,N+6},
\end{gathered}\end{equation}
where $N$ is the integer for one of the Fuchs indices.
So, in the case when \eqref{PT5} we obtain the following Fuchs indices
\begin{equation}\begin{gathered}
\label{PT4}
j_1=-1,\quad j_2=4,\quad \j_{3}=N,\quad j_4=-N-1.
\end{gathered}\end{equation}
We have not had any possibility to check the third step of the Painlev\'e test for all $N$. However we have found that equation \eqref{PDE} has the Painlev\'e property in the following cases : 1) $\alpha=0$, $\beta=0$, $\gamma=0$, $\delta \neq 0$; 2) $\delta=0$, $\alpha \neq 0$, $\beta \neq 0$ $\gamma \neq 0$; and 3) $\alpha=0$, $\beta \neq 0$, $\gamma=2\,\beta$. The fiirst and the second cases give the well-known  equations.

Taking into account these results let us try to obtain the Lax pair corresponding equation \eqref{PDE} in case of $\alpha=0$ and $\gamma=2\,\beta$.

\section{Lax pair for equation \eqref{PDE} at $\alpha=0$ and $\gamma=2\,\beta$}

Let us note that equation \eqref{PDE} has two partial cases. One of them  can be transformed to the well-known modified Korteweg-de Vries equation at $\delta=0$ and the other can be reduced at $\alpha=0$, $\beta=0$ and $\gamma=0$ to the Sine-Gordon equation. It is well known that the Cauchy problem for both the modified Korteweg-de Vries equation and the Sine-Gordon equation can be solved by the Inverse scattering transform. Moreover equation \eqref{PDE} is similar to the nonlinear integrable ordinary differential equation studied in works \cite{Kudr02a, Kudr02b}. Let us demonstrate that one can use the AKNS scheme to obtain the Lax pair for the partial case of equation
\eqref{PDE}. With this aim we look for the Lax pair in the form
\begin{equation}\begin{gathered}
\label{L1}
\hat{P}_t - \hat{Q}_x+[\hat{P}, \hat{Q}]=0,
\end{gathered}\end{equation}
where matrixes $\hat{P}$ and $\hat{Q}$ take the form
\begin{equation}
\label{L2}
\hat{P}=\begin{pmatrix}-i\,\lambda & & q\,\\
r & & i\,\lambda\end{pmatrix}, \qquad \hat{Q}=\begin{pmatrix} h & & e \\ f & & -h \end{pmatrix}.
\end{equation}

 We have the following equations for the matrix elements $\hat{Q}$ in the form:
\begin{equation}\begin{gathered}
\label{L3}
q_{t}-2\,h\,q-\,e_x+2\,i\,\lambda \,e=0, \\
\\
r_{t}+2\,h\,r-f_x-2\,i\,\lambda\,f=0,\\
\\
h_x-q\,f+e\,r=0.
\end{gathered}\end{equation}
Substituting expressions for $h(x,t)$, $f(x,t)$ and $e(x,t)$ in the form:
\begin{equation}\begin{gathered}
\label{L4}
h(x,t)=\frac{h_4}{\lambda}+h_3+h_2\,\lambda+h_1\,\lambda^2+h_0\,\lambda^3, \\
\\
f(x,t)=\frac{f_4}{\lambda}+f_3+f_2\,\lambda+f_1\,\lambda^2,\\
\\
e(x,t)=\frac{e_4}{\lambda}+e_3+e_2\,\lambda+e_1\,\lambda^2.
\end{gathered}\end{equation}
into equation \eqref{L3} we obtain at $r=\frac12\,u_x$ and $q=-\frac12\,u_x$ the following
Lax pair
\begin{equation}\begin{gathered}
\label{L6a}
\psi_{1,x}=-i\,\lambda\,\psi_1-\frac12\,u_x\,\psi_2,\\
\\
\psi_{2,x}=\frac12\,u_x\,\psi_1+i\,\lambda\,\psi_2,
\end{gathered}\end{equation}
\begin{equation}\begin{gathered}
\label{L6b}
\psi_{1,t}=h\,\psi_1+e\,\psi_2,\\
\\
\psi_{2,t}=f\,\psi_1-h\,\psi_2,
\end{gathered}\end{equation}
where elements $h$, $e$ and $f$ take the forms:
\begin{equation}\begin{gathered}
\label{L7}
h=-\frac{i\,\delta}{4\,\lambda}\,\cos{u}-i\,\beta\,u_x^2\,\lambda+8\,i\,\beta\,\lambda^3, \\
\\
f=-\frac{i\,\delta}{4\,\lambda}\,\sin{u}-\beta\,u_{xxx}-\frac{\beta}{2}\,u_{x}^3+2\,i\,\beta\,u_{xx}\,\lambda+4\,\beta\,u_x\,\lambda^2,\\
\\
e=-\frac{i\,\delta}{4\,\lambda}\,\sin{u}+\beta\,u_{xxx}+\frac{\beta}{2}\,u_{x}^3+2\,i\,\beta\,u_{xx}\,\lambda-4\,\beta\,u_x\,\lambda^2.
\end{gathered}\end{equation}

System of equations \eqref{L6a} and \eqref{L6b} can be used for the solution of the Cauchy problem for equation \eqref{PDE} at $\alpha=0$ and $\gamma=2\,\beta$.

\section{The Cauchy problem for equation \eqref{PDE} at $\alpha=0$ and $\gamma=2\,\beta$}

Let us illustrate how one can solve the Cauchy problem for equation \eqref{PDE} by means of the Inverse Scattering transform at $\alpha=0$ and $\gamma=2\,\beta$.
In this case the equation \eqref{PDE} takes the form
\begin{equation}\begin{gathered}
\label{T0}
u_{xt}+3\,\beta\,u_{x}^2\,u_{xx}+2\,\beta\,u_{xxxx}=\delta\,\sin {u}
\end{gathered}\end{equation}

The class of initial functions for the potential $u(x,t)$
of equation \eqref{T0} is determined by analogy with the initial function for
the Sine-Gordon  equation and takes the form
\begin{equation}\begin{gathered}
\label{T1}
\int_{-\infty}^{\infty} \,|u_x|\,dx < \infty, \qquad \lim_{x \to \pm \infty} u =\pi\,k.
\end{gathered}\end{equation}

Taking into account the Lax pair we get the following scalar functions \cite{Ablowitz, Drazin, Polyanin}
\begin{equation}
\label{T2}
\begin{pmatrix}\psi_1\,\\
\psi_2\end{pmatrix}  \rightarrow \begin{pmatrix}0\,\\
e^{-i\,\lambda\,x}\end{pmatrix}, \qquad x\rightarrow -\infty.
\end{equation}
and
\begin{equation}
\label{T3}
\begin{pmatrix}\psi_1\,\\
\psi_2\end{pmatrix}  \rightarrow \begin{pmatrix} b(\lambda,t)\,e^{i\,\lambda\,x}\,\\
 a(\lambda,t)\,e^{-i\,\lambda\,x}\end{pmatrix}, \qquad x\rightarrow +\infty.
\end{equation}
The properties of the scattering data are similar on data for the Sine-Gordon equation.

The Gelfand-Levitan-Marchenko equation for equation \eqref{T0} can be presented as the system of integral equations \cite{Ablowitz, Drazin}
\begin{equation}\begin{gathered}
\label{T4}
K_{11}(x,z,t)+\int_{x}^{\infty}\,K_{12}(x,y,t)\,f(y+z;t)\,dy,\\
\\
K_{12}(x,z,t)-f(x+z;t)-\int_{x}^{\infty}\,K_{11}(x,y;t)\,f(y+z;t)dy=0,
\end{gathered}\end{equation}
where
\begin{equation}\begin{gathered}
\label{T5}
f(X;t)=\frac{1}{2\,\pi}\,\int_{-\infty}^{\infty}\,\frac{b(k)}{a(k)}\,\exp{\left(ik\,X+\frac{i\,\delta\,t}{2\,k}-\,i\,\beta\,k^3\,t \right)}\,dk-\\
\\
-i\sum_{n=1}^{N}c_n\,\exp{\left\{i\,\zeta_n\,X+\frac{i\,\delta\,t}{2\,\zeta_{n}}-i\,\beta\,\zeta_n^3\,t\right\}}.
\end{gathered}\end{equation}
Here $b(k)$, $a(k)$ are the appropriate scattering data, and $c_n$, $n=1, \ldots N$, are the normalisation constants at
each discrete eigenvalue, $\zeta=\zeta_n$, all at $t=0$.

Solution of the Cauchy problem for equation \eqref{T0} is found by the formula \cite{Ablowitz, Drazin }
\begin{equation}\begin{gathered}
\label{T6}
u(x,t)=-2\, K_{12}(x,x,t).
\end{gathered}\end{equation}

$N$-soliton solutions correspond to reflectionless potentials when $b(k)=0$. In this case the system of equations \eqref{T4} is transformed to  the linear system of algebraic equations with the matrix
\begin{equation}\begin{gathered}
\label{T7}
A(x,t)=\{A_{kj(x,t)}\}, \qquad (k,\,\,j=1,1,\ldots N),
\end{gathered}\end{equation}
where
\begin{equation}\begin{gathered}
\label{T8}
A_{k \,j}(x,t)=\frac{\zeta_j}{\lambda_k+\lambda_j}\,\exp{\left(i\,\lambda_j\,x+\frac{i\,\delta\,t}{\lambda_j}-2\,\beta\,\lambda_j^3\,t\right)}.
\end{gathered}\end{equation}

$N$-soliton solution of equation \eqref{T0} can be found by formula
\begin{equation}\begin{gathered}
\label{T9}
u(x,t)=- \frac {i}{2}\, \ln \left\{\frac{\det(I+A(x,t))}{\det(I-A(x,t)}\right\},
\end{gathered}\end{equation}
where $I$ is the identity  $N \times N$ matrix, and elements of matrix $A(x,t)$ are determined by formula \eqref{T8}.
Let us study in more detail one and two-soliton solutions of equation \eqref{T0}.

\section{The one and two-soliton solutions of \eqref{T0} }

Using formula \eqref{T9} we can present the one-soliton solution of equation \eqref{T0} in the form:
\begin{equation}\begin{gathered}
\label{S1}
u(x,t)= \pm\, {2}\,{i}\,\ln\left\{\frac{1-i\,\exp{(\mu+\lambda\,x+\frac{\delta\,t}{\lambda}-2\,\beta\,\lambda^3\,t)}}
{1+i\,\exp{(\mu+\lambda\,x+\frac{\delta\,t}{\lambda}-2\,\beta\,\lambda^3\,t)}}\right\}.
\end{gathered}\end{equation}

This expression can be transformed to the one-soliton solution for equation \eqref{T0} in the form:
\begin{equation}\begin{gathered}
\label{o1}
u(x,t)=\pm\,4\,\arctan \left( {e}^{\theta} \right), \quad  \theta=\mu+\lambda\,x+\frac {\delta\,t}{
\lambda}-2\,\beta\,\lambda^{3}\,t
\end{gathered}\end{equation}

\begin{figure}
\center
\includegraphics[width=6.3cm,height=5.2cm]{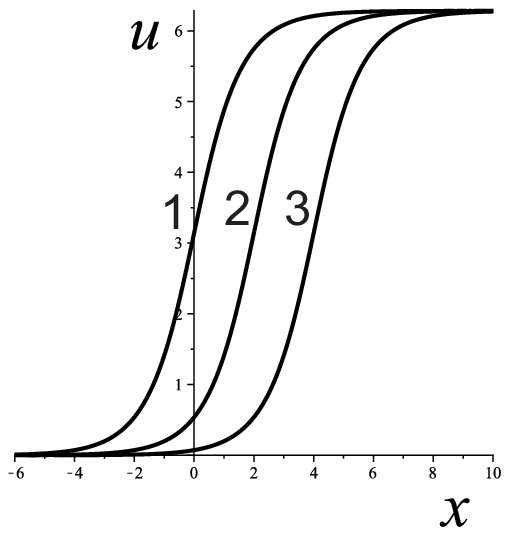}
\includegraphics[width=6.3cm,height=5.2cm]{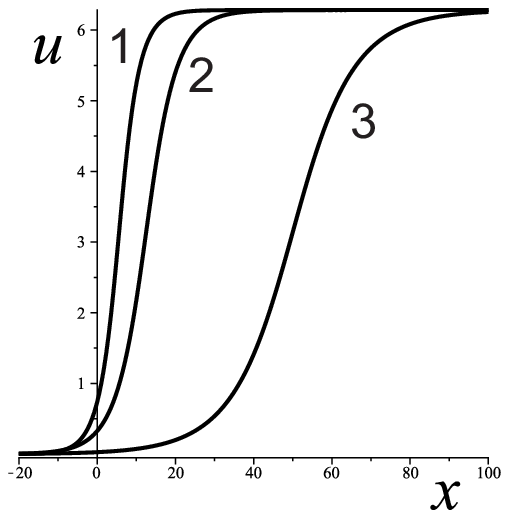}
\caption{The evolution of the one-soliton solution for equation \eqref{T0} at $\delta=-1$,  $\beta=0.5$ and $\lambda=1.0$ in time $t=0.0,\,\,1.0,\,\,2.0$ (left); The one-soliton solution of equation \eqref{T0} in time $t=0.5$ at $\beta=0.5$ for different values of $\lambda$: $\lambda=0.1,\,\,0.2,\,\,0.3$ (right).}
\label{FF1}
\end{figure}

The one-soliton solution is illustrated at $\delta=-1$ in Fig. \ref{FF1} for different values $\lambda $ and parameters $\mu=0.0$ and $\beta=0.1$ in time $t=1.0$. We have taken the following values of parameter $\lambda=0.1;\,\,0.2;\,\,0.4$ (left picture) and $\lambda=1.0;\,\,2.0;\,\,4.0$ (right picture).

The dependence on the velocity $\frac{dx}{dt}$ of the soliton is determined by formula
\begin{equation}\begin{gathered}
\label{o2}
\frac{dx}{dt}=-\frac{\delta}{\lambda^2}+2\,\beta\,\lambda^2
\end{gathered}\end{equation}

\begin{figure}
\center
\includegraphics[width=6.3cm,height=5.2cm]{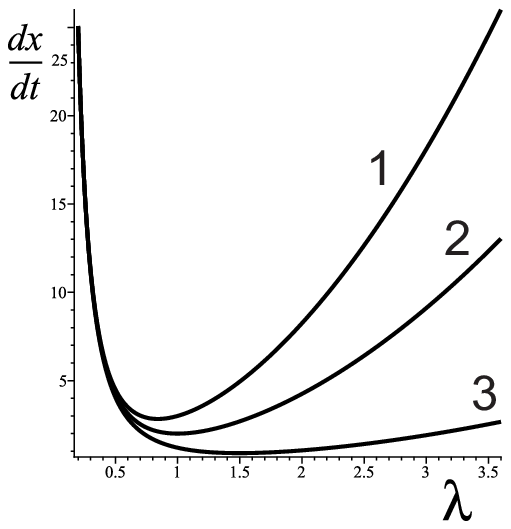}
\includegraphics[width=6.3cm,height=5.2cm]{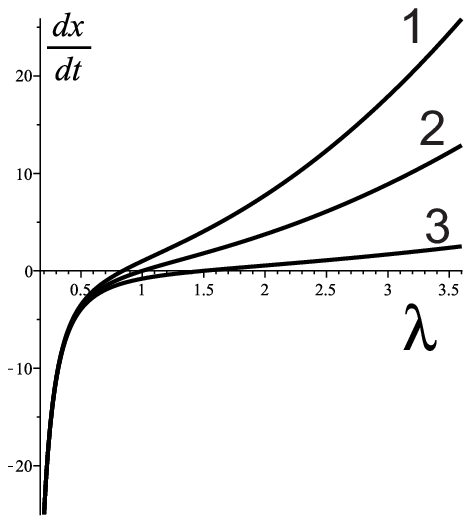}
\caption{The dependence of velocity for soliton \eqref{o1} on value $\lambda$ for different values $\beta=0.1;\,\,0.5; \,\,1.0$ at $\delta=-1$ (left) and $\delta=1$ (right).}
\label{FF2}
\end{figure}

\begin{figure}
\center
\includegraphics[width=6.3cm,height=5.2cm]{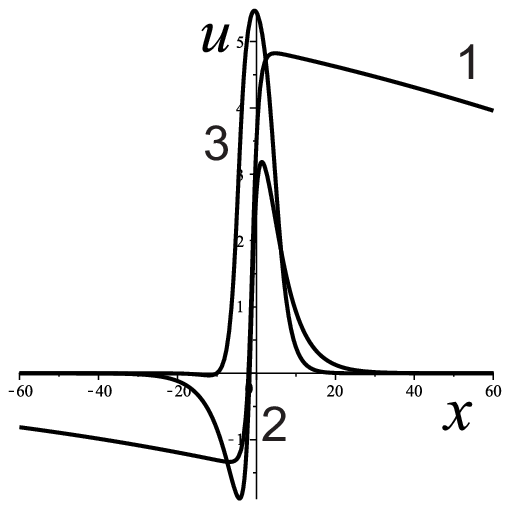}
\includegraphics[width=6.3cm,height=5.2cm]{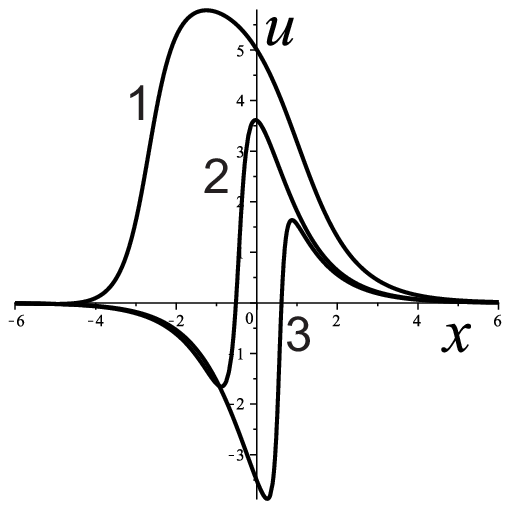}
\caption{The two-soliton solution of equation \eqref{T0} in time 0.01 at $\delta=-1$, $\mu_1=0.0$, $\mu_2=1.0$ and $\beta=0.5$ for different values of $\lambda$: curve line 1 at $(\lambda_1=0.01$ and $\lambda_2=0.99$); curve line 2 at $\lambda_1=0.2$ and $\lambda_2=0.8$; curve 3 at $\lambda_1=0.45$ and $\lambda_2=0.55$ (figure left); curve line 1 at $(\lambda_1=1.1$ and $\lambda_2=2.2$); curve line 2 at $\lambda_1=1.1$ and $\lambda_2=6.2$; curve 3 at $\lambda_1=1.1$ and $\lambda_2=10.2$ (figure right). }
\label{FF3}
\end{figure}

There is the minimum of velocity for one-soliton solution at $\delta=-1$. This minimum of the velocity depends on $\beta $ and can be found by formula
\begin{equation}\begin{gathered}
\label{o3}
\lambda_m=\beta^{-1/4}
\end{gathered}\end{equation}
The dependencies on velocity $\frac{dx}{dt}$ are illustrated in Fig.\ref{FF2} at $\delta=-1$ (left) and at $\delta=1$ (right). One can see on Figure in the left hand side that at $\lambda<\lambda_m$ the velocity of soliton decreases with increasing of $\lambda$ but at  $\lambda>\lambda_m$ vice versa.

The two-soliton solution can be written in the following form
\begin{equation}\begin{gathered}
\label{o4}
u(x,t)=\pm\,4\,\arctan \left( \left(\frac{\lambda_1+\lambda_2}{\lambda_1-\lambda_2}\right)\frac{e^{\theta_1}-e^{\theta_2}}{1+e^{\theta_1+\theta_2}} \right), \\
\\
\theta_i=\mu_i+\lambda_i\,x+\frac{\delta\,t}{\lambda_i}-2\,\beta\,\lambda_i^3\,t, \qquad (i=1,\,\,2).
\end{gathered}\end{equation}

\begin{figure}
\center
\includegraphics[width=6.3cm,height=5.2cm]{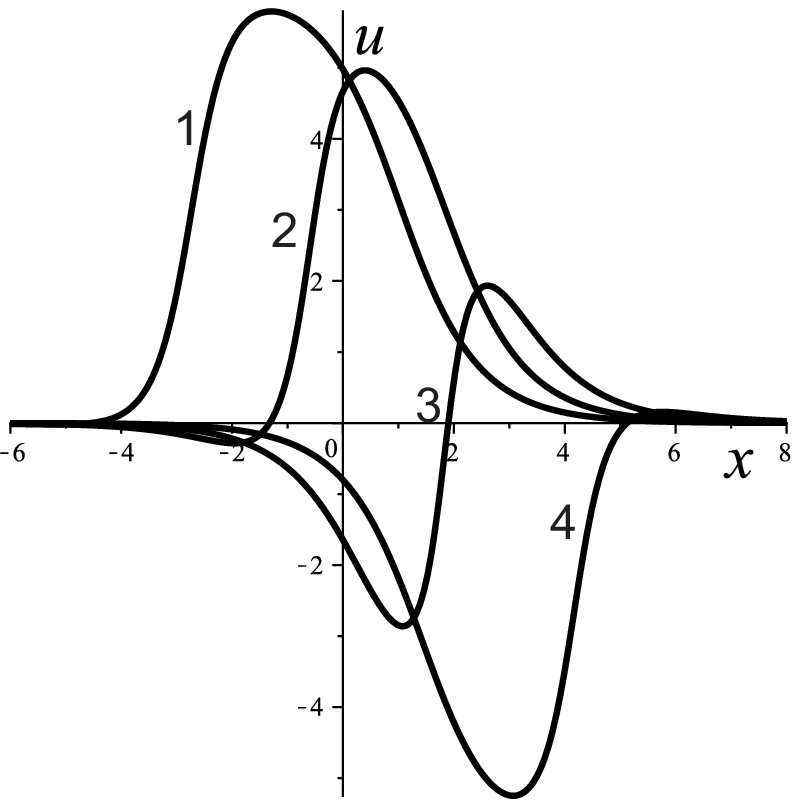}
\includegraphics[width=6.3cm,height=5.2cm]{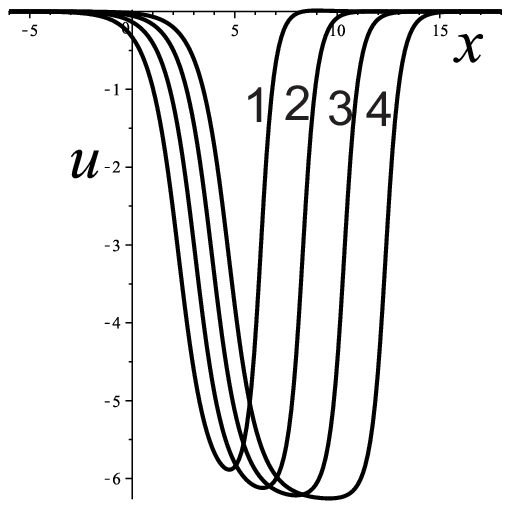}
\caption{The two-soliton solution of equation \eqref{T0} at $\delta=-1$, $\mu_1=0.0$, $\mu_2=5.0$, $\lambda_1=1.1$, $\lambda_2=2.2$ and $\beta=0.5$ for different values of time: $t=0.0,\,\,0.4,\,\,0.8,\,\,1,2$ (curve line 1 - 4 in left); $t=1.6,\,\,2.0,\,\,2.4,\,\,2.8$ (curve lines 1-4 in right).}
\label{FF4}
\end{figure}

The two-soliton solutions of equation \eqref{T0}  are illustrated in time $t=0.01$ in Fig. \eqref{FF3} for different values of lambda at $\mu_1=0.0$, $\mu_2=1.0$ and $\beta=0.5$. One can see that the form of two-soliton solutions depends on the values of parameters $\lambda_1$ and $\lambda_2$. For pictures on the left hand side we have taken the parameters  $ \lambda_1$ and $\lambda_2$ both less than 1: $\lambda_1=0.01$ and $\lambda_2=0.99$; $\lambda_1=0.2$ and $\lambda_2=0.8$; $\lambda_1=0.45$ and $\lambda_2=0.55$. In the right-hand side we have taken the values of parameter $\lambda$ more than 1: $\lambda_1=1.1$ and $\lambda_2=2.2$; $\lambda_1=1.1$ and $\lambda_2=6.2$; $\lambda_1=1.1$ and $\lambda_2=10.2$.

\begin{figure}
\center
\includegraphics[width=6.3cm,height=5.2cm]{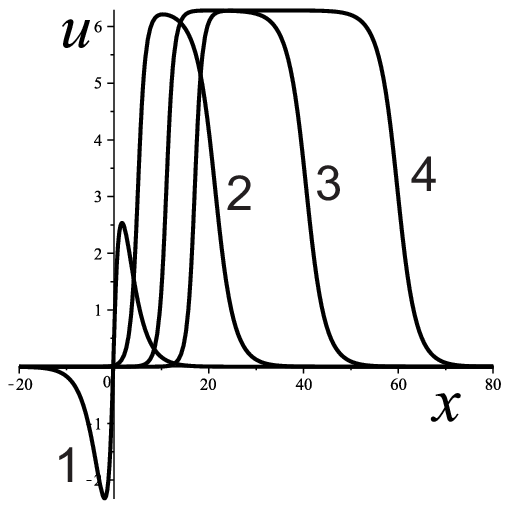}
\includegraphics[width=6.3cm,height=5.2cm]{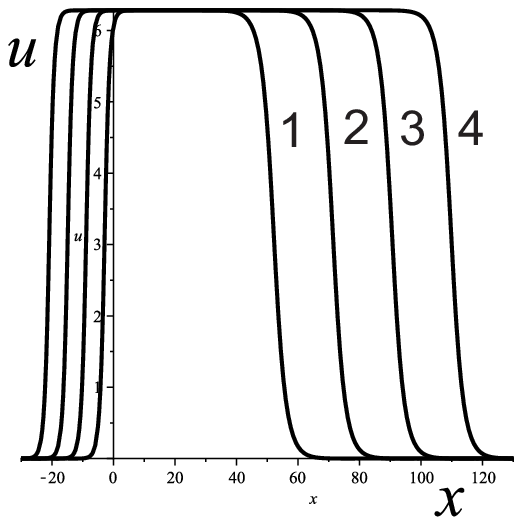}
\caption{The two-soliton solution of equation \eqref{T0} at $\delta=-1$, $\lambda_1=0.4$, $\lambda_2=1.0$ and $\beta=0.5$ for different values of time: $t=0.0,\,\,3.0,\,\,6.0,\,\,9.0$ for two values of $\mu_1=0.0,\,\,\mu_2=0.1$ (left) and $\mu_1=-20.0, \,\, \mu_2=20.0$ (right) .}
\label{FF5}
\end{figure}

The evolution of the two-soliton solution in time is demonstrated in Fig.\ref{FF4} at different value of time at $\mu_1=0$, $\mu_2=5.0$, $\beta=0.5$, $\lambda_1=1.1$ and $\lambda_2=2.2$. We have observed different moments of time and we can observe the change of two-soliton solution in time from $t=0.0$ to $t=2.8$.

\begin{figure}
\center
\includegraphics[width=6.3cm,height=5.2cm]{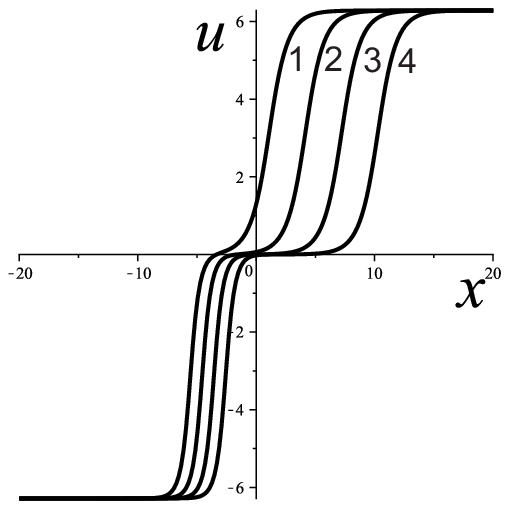}
\includegraphics[width=6.3cm,height=5.2cm]{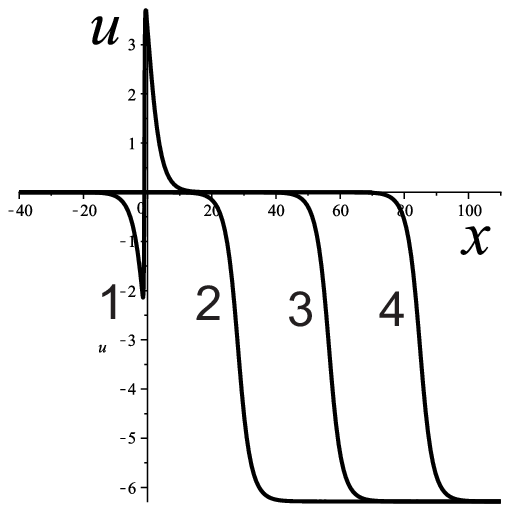}
\caption{The two-soliton solution of equation \eqref{T0} at $\delta=-1$, $\mu_1=0.0$, $\mu_2=10,0$ for different moments of time: $t=0.0,\,\,3.0,\,\,6.0,\,\,9.0$ and for different values of $\lambda_1$, $\lambda_2$ and $\beta$: $\beta=0.01$, $\lambda_1=-1$ and $\lambda_2=2.0$ (left) and $\beta=10.0$, $\lambda_1=0.4$ and $\lambda_2=10.0$ (right) .}
\label{FF6}
\end{figure}

The Influence of values $\mu_1$ and $\mu_2$ that determine the initial position of each of one soliton for the two-soliton solution is demonstrated in Fig.\ref{FF5}. We consider the following four moments of time $t=0.0,\,\,3.0,\,\,6.0,\,\,9.0$ at $\beta=0.5$, $\lambda_1=0.4$ and $\lambda=1.0$. One can see the changes of two soliton solution at the beginning of the time. However later we can observe more natural behaviour of solution.

The form of the two-soliton solution of equation \eqref{T0} can depend not only on the values of constants $\mu_1$, $\mu_2$, $\lambda_1$ and $\lambda_2$ but also on the value of the parameter $\beta$. These dependencies are illustrated in Fig. \ref{FF6} for two values of $\beta$. At first we took firstly the value $\beta=0.001$ and we could observe the two-soliton solution on the left-hand side. However if we take the value of parameter $\beta$ equal $10.0$ we obtain the behaviour of the two-soliton solution of equation \eqref{T0} on the right-hand side of Fig. \ref{FF6}.

We believe that equation \eqref{T0} is the first member of the following integrable hierarchy
\begin{equation}\begin{gathered}
\label{LL9}
u_{xt}+2\,\beta\,\frac{\partial}{\partial x}\left(u_x-i\,\frac{\partial}{\partial x}\right)L_n{\left[\frac{i\,u_{xx}}{2}+\frac{u_x^2}{4}\right]}=\delta\,\sin {u},
\end{gathered}\end{equation}
where $L_{n}[v]$ is the Lenard recursion operator which is determined by means of formula \cite{Lax}
\begin{equation}\begin{gathered}
\label{LL10}
\frac{L_{n+1}[v]}{\partial x}=\left(\frac{\partial^3}{\partial x^3}+4\,v\frac{\partial}{\partial x}+2\,\frac{\partial v}{\partial x}\right)\,L_{n}[v],\qquad L_{1}[v]=v.
\end{gathered}\end{equation}

We have checked that the equation
\begin{equation}\begin{gathered}
\label{LL11}
u_{xt}+\frac52\,u_{x}^2\,u_{xxxx}+\frac52\,u_{xx}^3+\frac{15}{8}\,u_{x}^4\,u_{xx}+\\
\\
+10\,u_x\,u_{xx}\,u_{xxx}+u_{xxxxxx}=\delta\,\sin u
\end{gathered}\end{equation}
of hierarchy \eqref{LL9} at $\beta=\frac12$ and $n=2$ has the one-soliton solution in the form
\begin{equation}\begin{gathered}
\label{LL12}
u=\pm 4\,\arctan\left\{\exp{\left(\mu+\lambda\,x+\frac{\delta\,t}{\lambda}-\lambda^5\,t\right)}\right\}.
\end{gathered}\end{equation}

We expect that the equations of hierarchy \eqref{LL9} are integrable. The Cauchy problem for equations of hierarchy \eqref{LL9} can be solved by the Inverse scattering transform. The one-soliton solution can be found taking into account the formula
\begin{equation}\begin{gathered}
\label{LL13}
u=\pm 4\,\arctan\left\{\exp{\left(\mu+\lambda\,x +\frac{\delta\,t}{\lambda}-2\,\beta\,\lambda^{2\,n+1}\,t\right)}\right\}.
\end{gathered}\end{equation}
So, we have studied equation \eqref{T0} which is the first member of integrable hierarchy \eqref{LL9}.

\section{Conclusion}

In this Letter we have studied the model that can be considered as the nonlinear interaction between dislocations in the rigid body. It turned out that our model is the generalization of two well-known dynamical systems of both the Frenkel-Kontorova system and the Fermi -Pasta-Ulam models. In fact we have considered the dynamical system that connects two well-known models. This new dynamical system has been studied taking into account the continuous limit approximation at $N \rightarrow \infty$ and $h \rightarrow 0$. In this case we have studied the dynamical system using the nonlinear partial differential equation. One of the main problems at the study of nonlinear partial differential equation is the question of the equation integrability. With this aim we have used the Painlev\'e test for studying integrability. We have obtained the Lax pair for the special case of an nonlinear  differential equation using the AKNS scheme. We have shown that the Cauchy problem for the special solution of the model for nonlinear dislocations can be solved by the Inverse Scattering transform. Both one and the two-soliton solutions were found. The behaviour of these soliton solutions for the obtained equation was studied at the different values of parameters.

\section*{Acknowledgment}

This research was supported by Russian Science Foundation grant No. 14-11-00258.

\end{document}